\documentclass[fleqn,usenatbib,letters]{mnras}

\usepackage{newtxtext,newtxmath}
\usepackage[T1]{fontenc}
\usepackage{ae,aecompl}

\usepackage{graphicx}	
\usepackage{amsmath}	
\usepackage{amssymb}	

\usepackage[percent]{overpic}
\usepackage[justification=centering, center]{caption}
\usepackage[flushleft]{threeparttable}
\usepackage{booktabs}
\usepackage{siunitx}

\usepackage{setspace}

\usepackage{adjustbox}

\usepackage{calligra}
\DeclareMathAlphabet{\mathcalligra}{T1}{calligra}{m}{n}
\DeclareFontShape{T1}{calligra}{m}{n}{<->s*[2.2]callig15}{}
\newcommand{\scriptr}{\mathcalligra{r}\,}

\usepackage[table,x11names]{xcolor}
\definecolor{aliceblue}{rgb}{0.94, 0.97, 1.0}

\usepackage[frozencache,cachedir=.]{minted} 
\usepackage[skins,minted]{tcolorbox}
\definecolor{mintedbackground}{rgb}{0.95,0.95,0.95}
\setminted[python]{
    bgcolor=mintedbackground,
    fontfamily=tt,
    linenos=false,
    numberblanklines=true,
    numbersep=12pt,
    numbersep=5pt,
    gobble=0,
    funcnamehighlighting=true,
    tabsize=4,
    obeytabs=false,
    mathescape=false,
    samepage=false,
    showspaces=false,
    showtabs =false,
    texcl=false,
    baselinestretch=1.1, 
    breaklines=true,
}

\newtcblisting{myminted}[2][]{listing engine=minted, listing only,#1, title=#2, minted language=python, colback=mintedbackground}
\newtcblisting{myminted2}[1][]{listing engine=minted, listing only,#1, minted language=python, colback=mintedbackground}

\title[A seismic scaling relation for stellar age II: The red giant branch]{A seismic scaling relation for stellar age II: The red giant branch}

\author[E.\ P.\ Bellinger]{Earl Patrick Bellinger$^{1,2}$\thanks{E-mail: bellinger@phys.au.dk}\\\vspace*{-2pt}%
$^{1}$SAC Postdoctoral Research Fellow, Stellar Astrophysics Centre, Department of Physics and Astronomy, Aarhus University, Denmark\\%
$^{2}$Visiting Fellow, School of Physics, University of New South Wales, Sydney, Australia\vspace*{-0.65\baselineskip}}

\usepackage{pifont}
\makeatletter
\newcommand\Pimathsymbol[3][\mathord]{%
  #1{\@Pimathsymbol{#2}{#3}}}
\def\@Pimathsymbol#1#2{\mathchoice
  {\@Pim@thsymbol{#1}{#2}\tf@size}
  {\@Pim@thsymbol{#1}{#2}\tf@size}
  {\@Pim@thsymbol{#1}{#2}\sf@size}
  {\@Pim@thsymbol{#1}{#2}\ssf@size}}
\def\@Pim@thsymbol#1#2#3{%
  \mbox{\fontsize{#3}{#3}\Piitsymbol{#1}{#2}}}
\newif\iftx@libertine
\newif\iftx@minion
\newif\iftx@coch
\newcommand{\Piitfont}[1]{\fontfamily{#1}\fontencoding{U}%
\fontseries{m}\fontshape{it}\selectfont}
\newcommand{\Piitsymbol}[2]{{\Piitfont{#1}\char#2}}
\makeatother
\newcommand{\newplus}{\Pimathsymbol[\mathbin]{ntxmia}{253}}

\date{Accepted XXX. Received YYY; in original form ZZZ}
\pubyear{2019}

\newif\ifref
\reftrue 
\reffalse
\definecolor{darkred}{rgb}{0.7, 0, 0}
\newcommand{\mb}[1]{\ifref\textcolor{darkred}{#1}\else #1\fi}

\begin{document}
\label{firstpage}
\pagerange{\pageref{firstpage}--\pageref{lastpage}}
\maketitle

\begin{abstract} 
    Owing to their simplicity and ease of application, seismic scaling relations are widely used to determine the properties of stars exhibiting solar-like oscillations, such as solar twins and red giants. 
    So far, no seismic scaling relations for determining the ages of red giant stars have been developed. 
    Such relations would be desirable for galactic archaeology, which uses stellar ages to map the history of the Milky Way. 
    The ages of red giants must instead be estimated with reference to grids of theoretical stellar models, which can be computationally intensive. 
    Here I present an exhaustive search for scaling age relations involving different combinations of observable quantities. 
    The candidate scaling relations are calibrated and tested using more than 1,000 red giant stars whose ages were obtained via grid-based modeling. 
    I report multiple high-quality scaling relations for red giant branch stars, the best of which are shown to be approximately as accurate as grid-based modeling with typical uncertainties of 15\%. 
    Additionally, I present new scaling mass and radius relations for red giants as well. 
\end{abstract} 



\section{Introduction} 
Cool stars like the Sun are observed to oscillate in a superposition of many small-amplitude radial and non-radial modes. 
The frequencies of these modes depend on the characteristics of the star, such as its mass and chemical composition \citep[e.g.,][]{2010aste.book.....A}. 
Consequently, asteroseismic measurements yield inferences on the global and internal properties of stars. 
The ages of these stars can furthermore be estimated by interpreting the oscillation data with reference to theoretical models of stellar evolution, assuming the models are approximately correct \citep[e.g.,][]{basuchaplin}. 

Stars continue to exhibit solar-like oscillations even after leaving the main sequence and ascending the red giant branch \citep[RGB; e.g.,][]{2017A&ARv..25....1H}. 
Thanks to the space-based asteroseismic surveys such as CoRoT \citep{2003AdSpR..31..345B}, \emph{Kepler} \citep{2010Sci...327..977B}, K2 \citep{2014PASP..126..398H}, and now TESS \citep{2015JATIS...1a4003R}, high-precision seismic data have been delivered for many thousands of giant stars \citep[e.g.,][]{2018ApJS..236...42Y}. 
Asteroseismic age dating of these stars has given birth to the field of galactic archaeology, which aims to uncover the history of the Milky Way by study of its stellar constituents \citep[e.g.,][]{2013MNRAS.429..423M}. 
The forthcoming PLATO \citep{2017AN....338..644M} and WFIRST \citep{2015JKAS...48...93G} missions furthermore promise to deliver orders-of-magnitude more asteroseismic observations, creating the need for fast and robust methods for processing the anticipated data yield. 

The interpretation of asteroseismic data has been greatly aided by the use of so-called \emph{scaling relations} \citep[e.g.,][]{1986ApJ...306L..37U, 1991ApJ...368..599B, 1995A&A...293...87K, Stello_2008, 2016MNRAS.460.4277G}. 
These relations estimate the properties of a star, such as its mass or radius, by scaling from reference values, which are generally chosen to be properties of the Sun. 
Scaling relations are widely used because they are accurate and precise \citep[e.g.,][]{2019ApJ...885..166Z} while mitigating the need for complex algorithms for interpreting asteroseismic data with reference to grids of theoretical stellar models. 
Given that red giants follow an age--mass relation \citep[e.g.,][]{2012ASSP...26...11M}, it follows that a scaling relation for red giant ages should exist as well. 
In a previous paper I have introduced a scaling relation for estimating the ages of main-sequence stars \citep[][hereinafter Paper I]{2019MNRAS.486.4612B}. 
In this work I aim to develop a similar relation for red giant stars. 

The reason for separate scaling relations for stellar age is that the seismic signature of age is different for stars on the main sequence and on the RGB. 
The age of a main-sequence star is linked to the amount of hydrogen in its core, which is probed by the small frequency separation. 
Stars on the RGB are void of hydrogen in their cores and undergo no fusion there; thus, the small separation is not useful for measuring their ages. 

\section{Data}
\begin{figure}%
    \centering%
    \includegraphics[width=\linewidth]{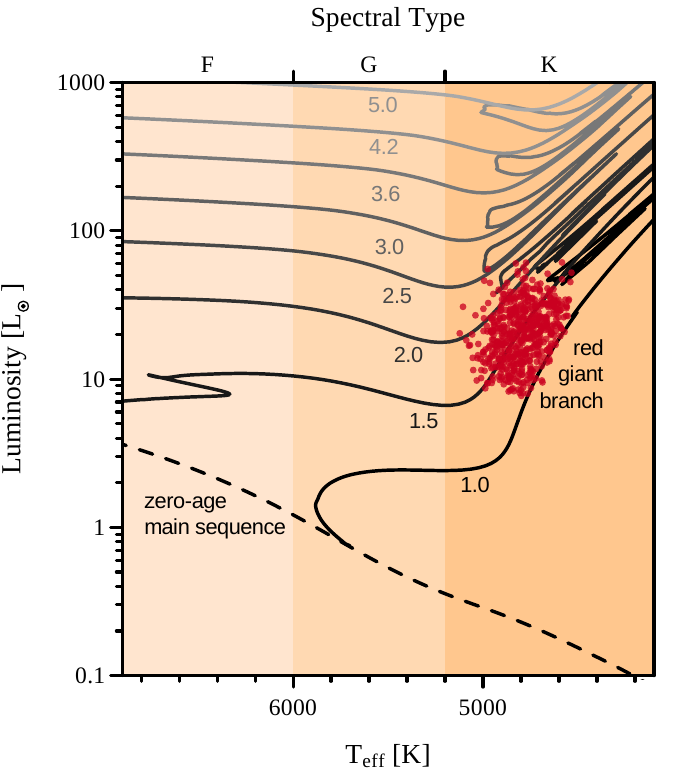}%
    \caption{Locations of the red giant stars used in this study in the HR diagram. 
    The background lines are MIST stellar evolution tracks of the indicated masses \citep{2016ApJ...823..102C}. \label{fig:hr}}%
\end{figure}

The data set for this study consists of 1,143 red giant stars that were observed during the nominal \emph{Kepler} mission and spectroscopically observed by APOGEE \citep{2009CoAst.160...74H, 2017AJ....154...94M, 2018ApJS..239...32P}. 
It represents a gold sample of stars whose measurements were of high enough quality to extract $g$-mode period spacings \citep{2016A&A...588A..87V}. 
The ages, masses and radii of these stars were estimated using grid-based modeling with BASTA \citep[BAyesian STellar Algorithm;][]{2015MNRAS.452.2127S, 2017ApJ...835..173S, 2018MNRAS.475.5487S}. 
Their positions in the Hertzsprung--Russell diagram are shown in Figure~\ref{fig:hr}. 
The parameter ranges of the stars are tabulated in Table~\ref{tab:ranges}. 

In order to account for potential systematic errors in the metallicity measurements, \mb{which are reported with uncertainties smaller than $0.05$~dex}, I have perturbed all [Fe/H] measurements by adding Gaussian noise with a standard deviation of $0.05$~dex. 
This did not make a substantial difference to the final results. 

Note that all of the stars considered here are on the RGB, i.e., none are clump stars. 
Thus, stars should first be classified as being RGB stars \citep[e.g.,][]{2019MNRAS.489.4641E} with values within the ranges of Table~\ref{tab:ranges} before using the presented relations. 

\begin{table}
    \centering
    \begin{threeparttable}
        \caption{Parameter Ranges for the Training Data \label{tab:ranges}}
        \begin{tabular}{ccccccc} \cline{1-7}\cline{1-7}
            {Parameter} & {Min.} & {Median} & {Max.} & {Solar} & {Unit} & {Source} \\\hline\hline
            $\nu_{\max}$ & 27.9 & 104.5 & 255.6 & 3090$^\text{(e)}$ & $\mu$Hz & (a) \\
            $\Delta\nu$ & 3.73 & 9.25 & 17.90 & 135.1$^\text{(e)}$ & $\mu$Hz & (a) \\
            $\Delta\Pi$ & 49.7 & 75.1 & 147.4 & -- & s & (b) \\
            $T_{\text{eff}}$ & 4520 & 4790 & 5120 & 5772$^\text{(f)}$ & K & (c) \\
            $\text{[Fe/H]}$ & -1.55 & 0.02 & 0.50 & 0 & -- & (c) \\\hline
            Mass & 0.96 & 1.33 & 2.58 & 1 & M$_\odot$ & (d) \\
            Radius & 3.95 & 6.60 & 14.24 & 1 & R$_\odot$ & (d) \\
            Age & 0.6 & 4.3 & 11.6 & 4.569$^\text{(g)}$ & Gyr & (d)  \\\cline{1-7}\cline{1-7}
        \end{tabular}
        \begin{tablenotes}[flushleft]
            \item\hspace{-0.8mm}\emph{Notes}. 
            The number of significant figures reflect the typical precision of each given measurement type. 
            Solar values are given for comparison. 
            References: \mb{
            $^\text{(a)}$\citet{2009CoAst.160...74H}; 
            $^\text{(b)}$\citet{2016A&A...588A..87V}; 
            $^\text{(c)}$\citet{2017AJ....154...94M}; 
            $^\text{(d)}$\citet{2018MNRAS.475.5487S}; 
            $^\text{(e)}$\citet{2011ApJ...743..143H}; 
            $^\text{(f)}$\citet{2016AJ....152...41P}; 
            $^\text{(g)}$\citet{2015A&A...580A.130B}.}
        \end{tablenotes}
    \end{threeparttable}
\end{table}

\section{Methods} \label{sec:methods}
I consider scaling relations for stellar age of the form
\begin{align}
    \frac{\tau}{\tau_\scriptr}
    &=
    \bigg(
        \frac{\nu_{\max}}{\nu_{\max,\scriptr}}
    \bigg)^\alpha
    \bigg(
        \frac{\Delta\nu}{\Delta\nu_\scriptr}
    \bigg)^\beta
    \bigg(
        \frac{T_{\text{eff}}}{T_{\text{eff},\scriptr}}
    \bigg)^\gamma
    \exp\bigg(
        \text{[Fe/H]}
    \bigg)^\delta \label{eq:scalingX}
\end{align}
where $\tau$ refers to the age of the star, $\nu_{\max}$ is the frequency at maximum power, $\Delta\nu$ is the large frequency separation, $T_{\text{eff}}$ is the effective temperature, and [Fe/H] is the metallicity. 
Quantities subscripted with $\scriptr$ are reference values used for nondimensionalization, which in each case I take to be the median of the observed values (see Table~\ref{tab:ranges}). 
Though it is customary to use solar values as reference, there is no reason why the ages of red giant stars should be tied to the age of the Sun. 
Note that [Fe/H] requires no reference value. 

The task is now to determine the optimal exponents ${\mathbf P = [\alpha, \beta, \gamma, \delta]}$ for predicting $\tau$, and also to consider relations in which subsets of these exponents are zero. 
By log-transforming both sides, this becomes an ordinary least squares problem without an intercept term. 
However, since the logarithms of $\Delta\nu$ and $\nu_{\max}$ are collinear \citep{2009MNRAS.400L..80S}, the stability of the solution is improved when a penalty is applied to the magnitude of the solution \citep{hoerl1970}. 
This is the ridge regression problem of \citet{Tikhonov}. 
Thus to find the optimal $\mathbf P$ we minimize 
\begin{equation}
    \mathcal{F}(\mathbf P) = \sum_i \left(
        \frac{\hat \tau_i(\mathbf P) - \tau_i}{\sigma_i}
    \right)^2 \newplus \lambda \sum_j P_j^2
\end{equation}
where $\hat\tau(\mathbf P)$ is the scaling value from Equation~\ref{eq:scalingX} for a given set of exponents $\mathbf P$; $\tau_i$ is the age of the $i$th star as estimated via grid-based modeling; $\sigma_i$ is the uncertainty on that estimate; and $\lambda$ is the ridge penalty, the value of which is chosen by 10-fold cross validation \citep[e.g.,][]{friedman2001elements}. 
Note that only the uncertainties in age are considered, because despite the resulting attenuation of $\mathbf P$, this gives the best exponents for predicting ages \citep[e.g.,][]{ammann1989standard}. 
This procedure is then repeated for the 15 nonempty subsets \mb{(combinations of parameters)} of $\mathbf P$. 
Scaling relations involving the period spacing $\Delta\Pi$ were also tested, but they were not found to improve the estimates. 
Finally, scaling relations for mass and radius are also sought. 

\section{Results \& Conclusions} 
\begin{figure*}
    \centering
    \adjustbox{trim={0 0.38in 0 0}, clip}{\includegraphics[width=0.5\linewidth]{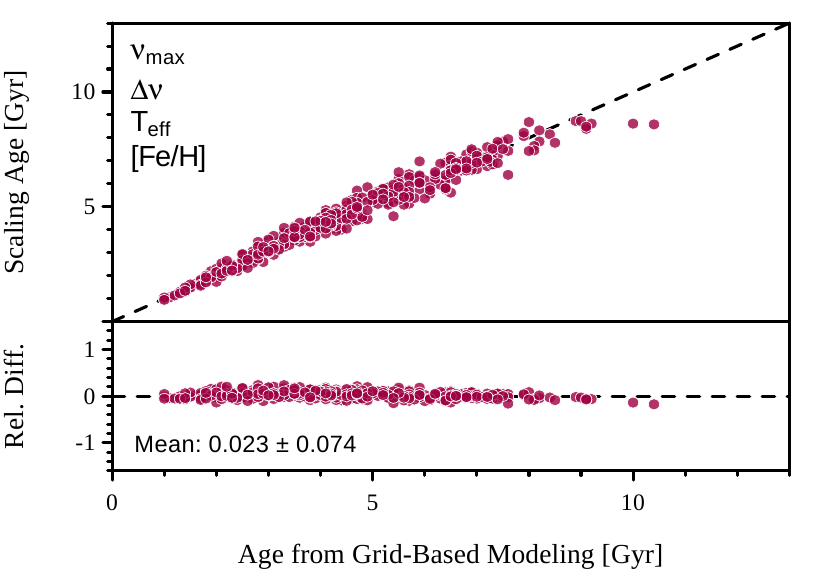}}%
    \adjustbox{trim={0.42in 0.38in 0 0}, clip}{\includegraphics[width=0.5\linewidth]{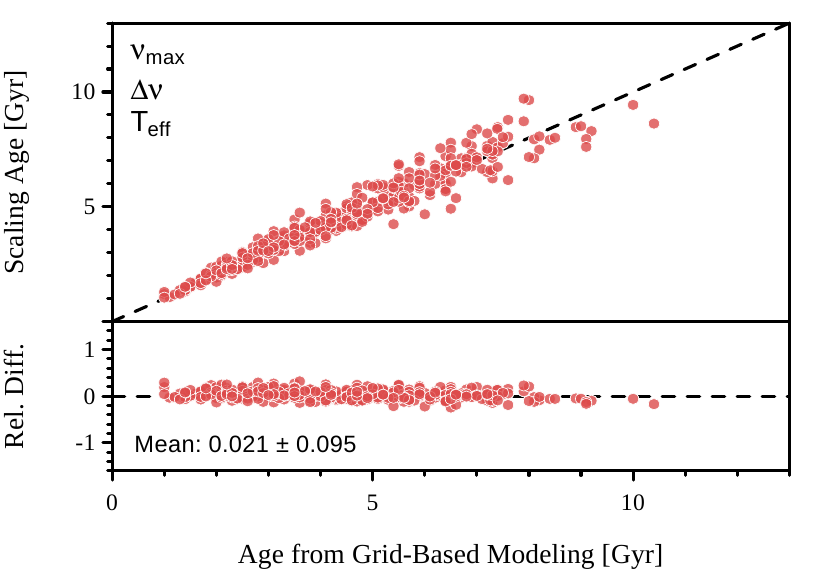}}\\
    \adjustbox{trim={0 0.38in 0 0}, clip}{\includegraphics[width=0.5\linewidth]{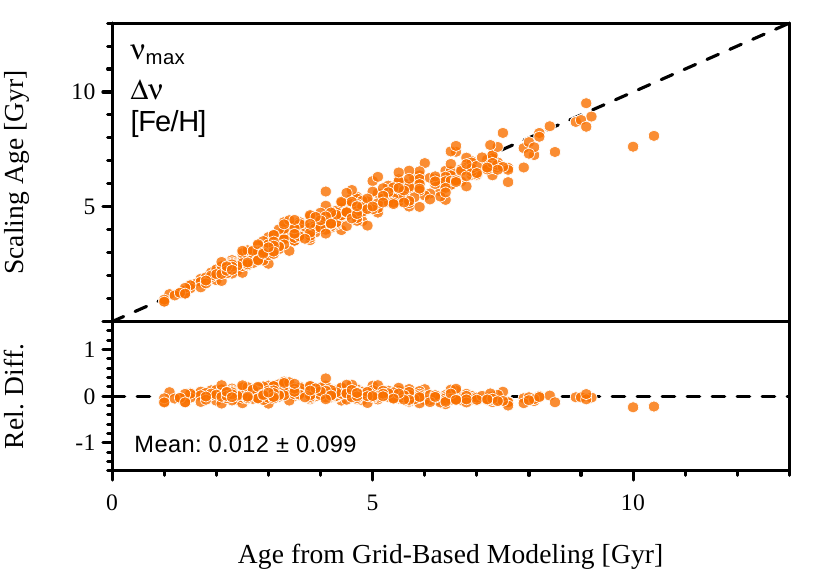}}%
    \adjustbox{trim={0.42in 0.38in 0 0}, clip}{\includegraphics[width=0.5\linewidth]{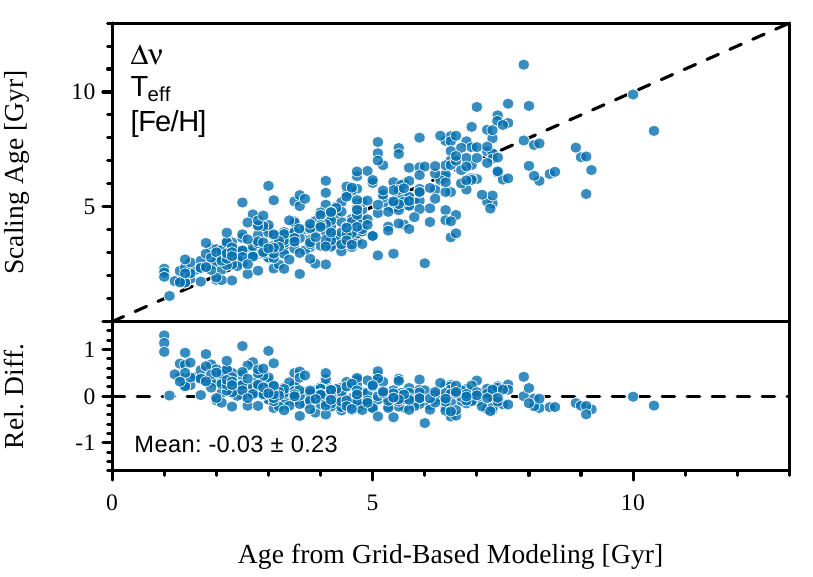}}\\
    \adjustbox{trim={0 0.38in 0 0}, clip}{\includegraphics[width=0.5\linewidth]{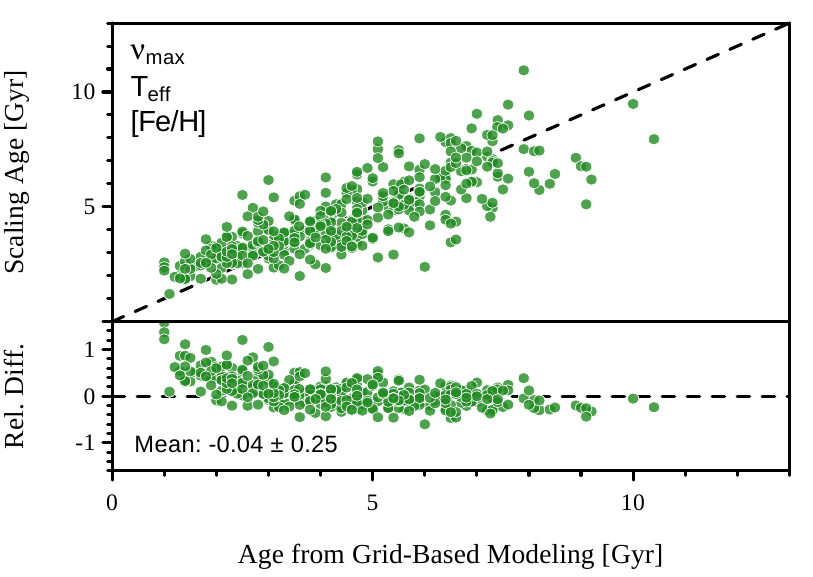}}%
    \adjustbox{trim={0.42in 0.38in 0 0}, clip}{\includegraphics[width=0.5\linewidth]{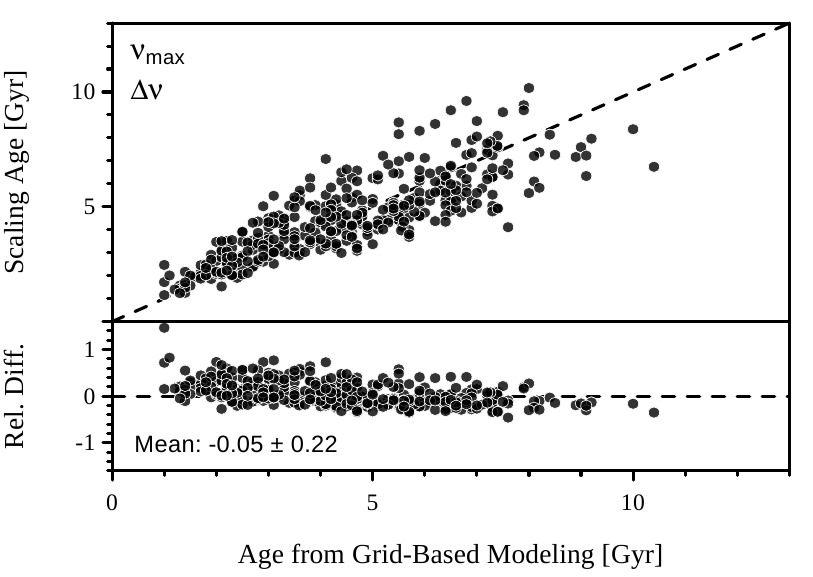}}\\
    \adjustbox{trim={0 0 0 0}, clip}{\includegraphics[width=0.5\linewidth]{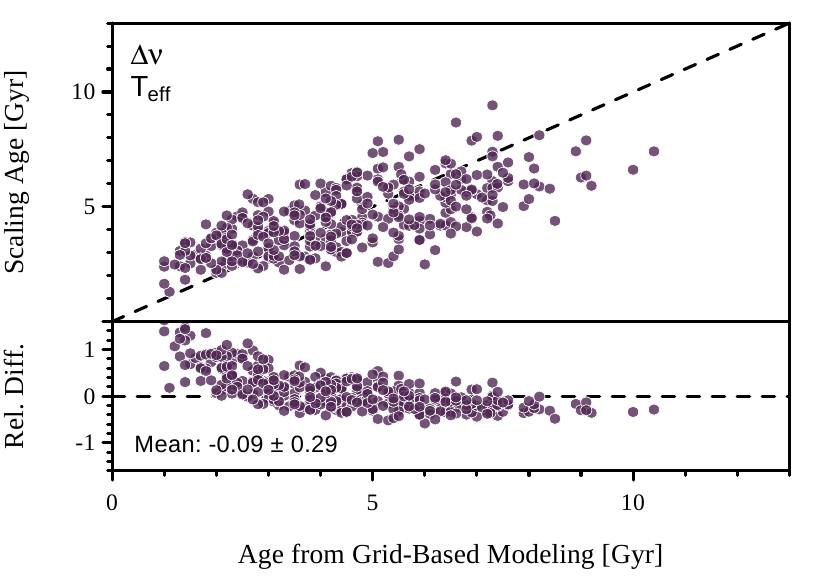}}%
    \adjustbox{trim={0.42in 0 0 0}, clip}{\includegraphics[width=0.5\linewidth]{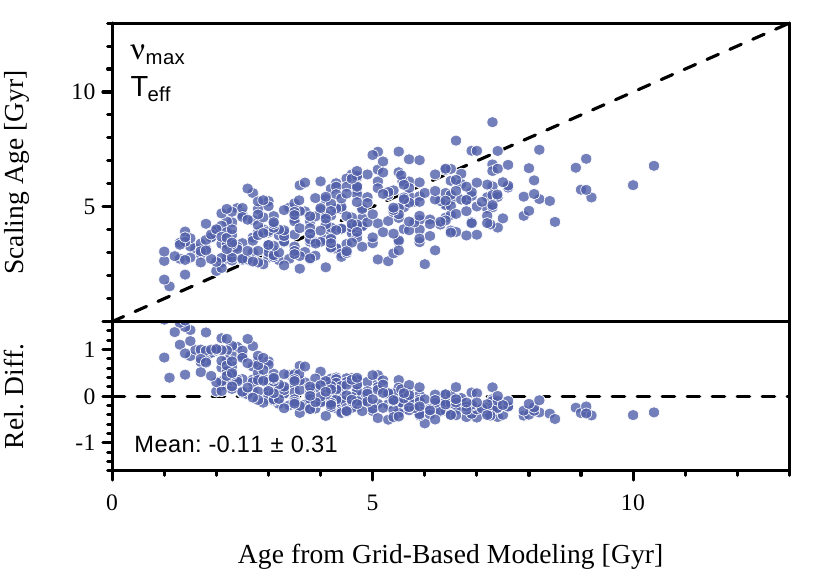}}
    \caption{Ages from calibrated scaling relations compared to the ages from grid-based modeling for 1,143 \emph{Kepler} red giants using the eight relations shown in Table~\ref{tab:exponents-age}. 
    The variable set is indicated in each panel. 
    Every scaling age estimate is made using 10-fold cross validation, i.e., 10\% of the data are held out, the relation is calibrated on the other 90\%, the relation is applied to the 10\% of held out data to estimate their ages (which are then plotted), and this is repeated for all folds of 10\%. 
        \label{fig:scaling-age}}
\end{figure*}

The discovered exponents for the scaling age relations are listed in Table~\ref{tab:exponents-age}. 
The quality of each relation can be assessed via the cross-validated coefficient of determination $r^2$ between the scaling and grid-based age. 
The best relation, which uses the full variable set (combination~1, ${r^2 = 0.99}$), is compared to the ages from grid-based modeling in the top left panel of Figure~\ref{fig:scaling-age}. 
This relation is approximately as accurate as grid-based modeling, with differences of only about 7\% on average, which is smaller than typical uncertainty. 

\begin{table*}
    \centering
    \begin{threeparttable}
        \caption{Calibrated Exponents for Red Giant Scaling Age Relations \label{tab:exponents-age}} 
        \begin{tabular}{S[table-format=1]
                      | S[table-format=3.4] 
                        S[table-format=3.4] 
                        S[table-format=3.4] 
                        S[table-format=3.4] 
                      | S[table-format=1.2]
                      | S[separate-uncertainty = true, table-format=2.3(3), table-align-uncertainty = true] 
                        S[table-format=1.2]}
\cline{1-8}\cline{1-8} & \textcolor{gray}{$\nu_{\max}$} & \textcolor{gray}{$\Delta\nu$} & \textcolor{gray}{$T_{\text{eff}}$} & \textcolor{gray}{$\text{[Fe/H]}$} &  & &    \\ \hline
{Combination} & {$\alpha$} & {$\beta$} & {$\gamma$} & {$\delta$} & {$\sigma_{\text{sys}}/$Gyr} & {Relative difference} & {$r^2$}  \\ \hline \hline
\rowcolor{aliceblue} 1   &       -9.760   &       13.08   &       -6.931          &   0.4894   &       0.25    &       0.023(74)       &       0.99 \\\hline
\rowcolor{aliceblue} 2   &       -7.778          &       10.77   &       -11.05      &        {--}    &       0.32    &       0.021(95)       &       0.98 \\
\rowcolor{aliceblue} 3   &       -12.19          &       15.86   &       {--}    &   1.027    &       0.34    &       0.012(99)       &       0.97 \\
4        &       {--}    &       1.396   &       -22.32          &       -1.046      &        0.82    &       -0.03(23)       &       0.83 \\
5        &       1.084   &       {--}    &       -23.28          &       -1.165      &        0.92    &       -0.04(25)       &       0.79 \\\hline
6        &       -8.837          &       11.73   &       {--}    &       {--}    &   0.86     &       -0.05(22)       &       0.86 \\
7        &       {--}    &       0.9727          &       -14.64          &       {--}         &       1.2     &       -0.09(29)       &       0.71 \\
8        &       0.6424          &       {--}    &       -13.82          &       {--}         &       1.3     &       -0.11(31)       &       0.64 \\
\cline{1-8}\cline{1-8}
        \end{tabular}
        \begin{tablenotes}
            \item\hspace{-0.8mm}\emph{Notes}. 
            Each row refers to a different combination of variables for Equation~\ref{eq:scalingX}. 
            The bases of the exponents are shown in gray at the top of the table for reference. 
            The rows are grouped by the number of variables. 
            Only variable combinations with ${r^2>0.4}$ are shown. 
            Rows with ${r^2>0.9}$, representing the best relations, are shaded blue. 
            Average relative differences between the scaling relation  estimates and the grid-based modeling estimates are given (\emph{cf.}~Figure~\ref{fig:scaling-age}). 
        \end{tablenotes}
    \end{threeparttable}
\end{table*}

The purely seismic scaling age relation (combination~6) works to some extent, giving ages which are on average within 22\% of the age from grid-based modeling. 
This opens the door to doing galactic archaeology from large-scale photometry missions even when spectroscopic follow-up is slow or infeasible. 
Relations with only $\nu_{\max}$ and spectroscopy (combinations~5 and 8) yield ages that are within 31\% of the modeling age, which permits age estimates for stars measured with low frequency resolution. 
However, it must be cautioned that these two relations tend to overestimate the ages of young stars. 

\begin{figure}
    \centering
    \includegraphics[width=\linewidth]{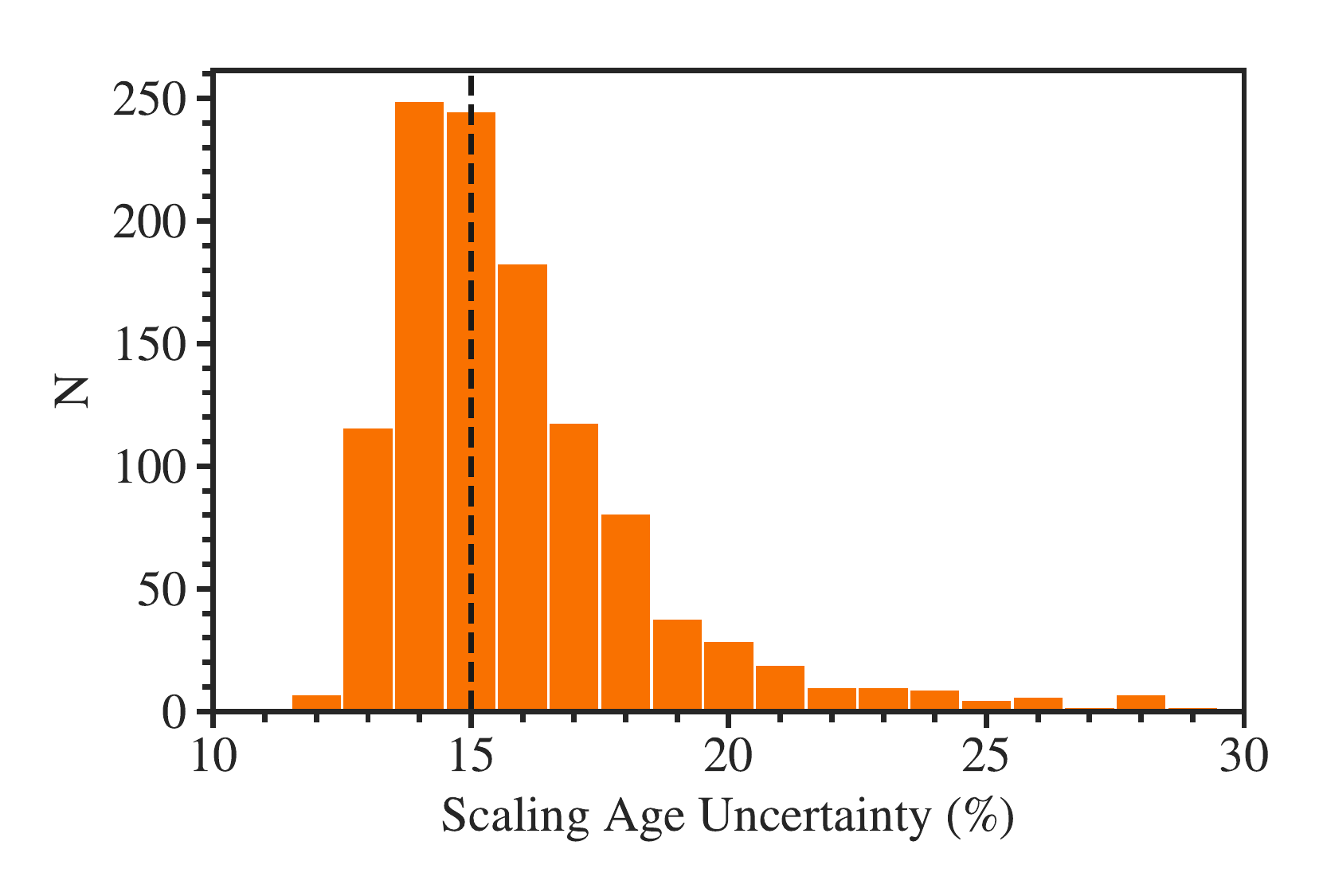}
    \caption{Histogram of relative age uncertainties from application of the full variable set scaling age relation to the selected \emph{Kepler} red giants. 
    The median uncertainty of 15\% is indicated. 
    } 
    \label{fig:uncertainties}
\end{figure}

An important question is one of the systematic errors that arise from use of these relations. 
Although it would be possible to report uncertainties in the estimated exponents, for example from bootstrapping, the values would be somewhat meaningless because the ridge regressor is a biased estimator. 
Instead, Table~\ref{tab:exponents-age} provides the root mean square error for each relation. 
This value, denoted $\sigma_{\text{sys}}$, should be added in quadrature to the random uncertainties obtained when applying each relation. 
Application of relation~1 gives a typical uncertainty of about 15\% (see Figure~\ref{fig:uncertainties}). 

\begin{figure*}
    \centering
    \adjustbox{trim={0 0.38in 0 0}, clip}{\includegraphics[width=0.5\linewidth]{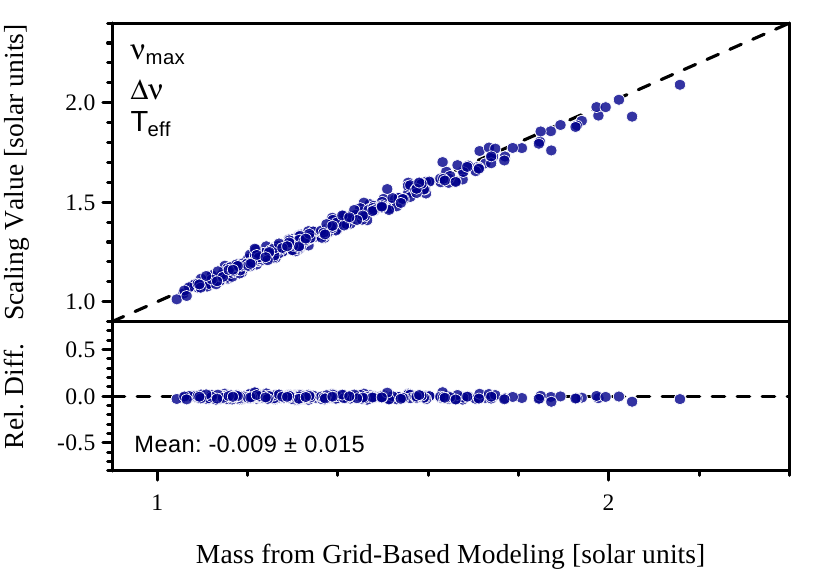}}%
    \adjustbox{trim={0.25in 0.38in 0 0}, clip}{\includegraphics[width=0.5\linewidth]{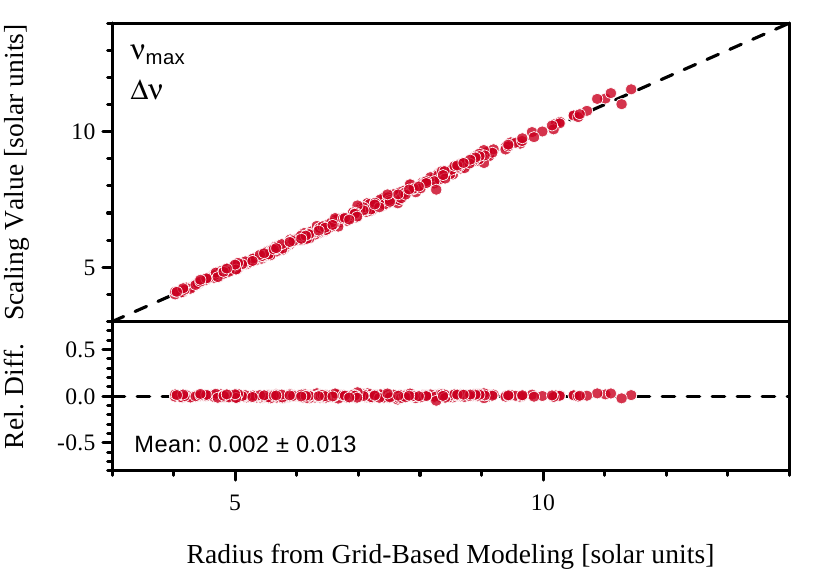}}\\
    \adjustbox{trim={0 0 0 0}, clip}{\includegraphics[width=0.5\linewidth]{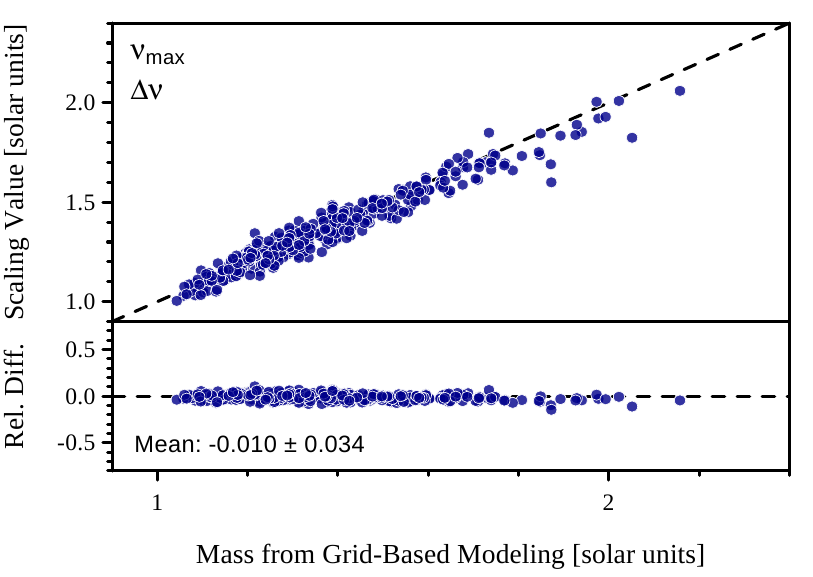}}%
    \adjustbox{trim={0.25in 0 0 0}, clip}{\includegraphics[width=0.5\linewidth]{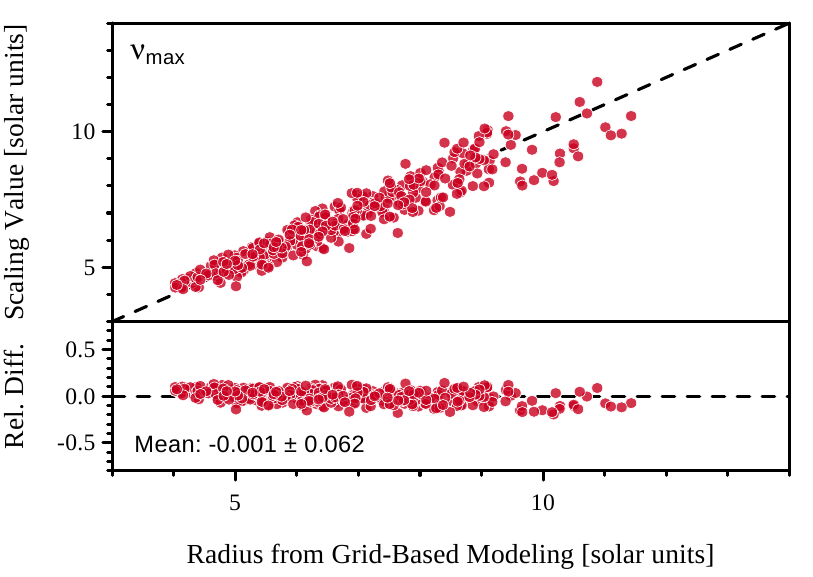}}
    \caption{Masses (left) and radii (right) from selected calibrated scaling relations (combinations 2, 6, and 10) compared to estimates from grid-based modeling. \vspace*{-0.2cm}
        \label{fig:MandR}}
\end{figure*}

Calibrated scaling relations for mass and radius are shown in Tables~\ref{tab:exponents-M} and \ref{tab:exponents-R}. 
Note that as discussed in Section~\ref{sec:methods} these calibrated relations use the median values for reference rather than solar values. 
Generally the classical scaling relations are recovered with small corrections as expected and work extremely well (${r^2 \sim 1}$). 
Also, as $T_\text{eff}$ changes relatively little on the RGB (at most $7\%$ from the median \mb{in this sample of low-luminosity RGB stars}), the purely-seismic relations ${M \propto \nu_{\max}^3\, \Delta\nu^{-4}}$ and ${R\propto \nu_{\max}\, \Delta\nu^{-2}}$ work to fairly high accuracy, differing from grid-based modeling estimates on average only by about $3\%$ and $1\%$, respectively. 
Similarly, ${R\propto \Delta\nu^{-0.7}}$ and ${R\propto 1/\sqrt{\nu_{\max}}}$ work to about $4\%$ and $6\%$. 
Some of these relations are visualized in Figure~\ref{fig:MandR}.

Naturally, these relations are only as good as the estimates to which they are fit. 
Different assumptions in the stellar models used to make those estimates would yield different scaling relations.

\begin{table*}
    \centering
    \begin{threeparttable}
        \caption{Calibrated Exponents for Red Giant Scaling Mass Relations \label{tab:exponents-M}} 
        \begin{tabular}{S[table-format=1]
                      | S[table-format=2.4] 
                        S[table-format=2.4] 
                        S[table-format=2.4] 
                        S[table-format=2.4] 
                      | S[table-format=1.3]
                      | S[separate-uncertainty = true, table-format=2.4(4), table-align-uncertainty = true] 
                        S[table-format=1.2]}
\cline{1-8}\cline{1-8} & \textcolor{gray}{$\nu_{\max}$} & \textcolor{gray}{$\Delta\nu$} & \textcolor{gray}{$T_{\text{eff}}$} & \textcolor{gray}{$\text{[Fe/H]}$} &  & &    \\ \hline
{Combination} & {$\alpha$} & {$\beta$} & {$\gamma$} & {$\delta$} & {$\sigma_{\text{sys}}/$M$_\odot$} & {Relative difference} & {$r^2$}  \\ \hline \hline
\rowcolor{aliceblue} 2   &       2.901   &       -3.876          &       1.621   &   {--}     &       0.023   &       -0.009(15)      &       0.98 \\
\rowcolor{aliceblue} 3   &       3.546   &       -4.619          &       {--}    &   -0.1457          &       0.023   &       -0.008(16)      &       0.98 \\
4        &       {--}    &       -0.3845         &       5.740    &       0.4290   &   0.10      &       -0.022(72)      &       0.70 \\
5        &       -0.2976         &       {--}    &       5.935   &       0.4594      &        0.11    &       -0.024(80)      &       0.65 \\\hline
\rowcolor{aliceblue} 6   &       3.056   &       -4.015          &       {--}    &   {--}     &       0.046   &       -0.010(34)      &       0.91 \\
\cline{1-8}\cline{1-8}
        \end{tabular}
        \begin{tablenotes}
            \item\hspace{-0.8mm}\emph{Notes}. Combination numbers are the same as those from Table~\ref{tab:exponents-age}. Combinations in which any of the variables are negligible, such as $\delta$ in combination~1, are omitted. 
        \end{tablenotes}
    \end{threeparttable}
\end{table*}

\begin{table*}
    \centering
    \begin{threeparttable}
        \caption{Calibrated Exponents for Red Giant Scaling Radius Relations \label{tab:exponents-R}} 
        \begin{tabular}{S[table-format=2]
                      | S[table-format=2.4] 
                        S[table-format=2.4] 
                        S[table-format=2.4] 
                        S[table-format=2.4] 
                      | S[table-format=1.3]
                      | S[separate-uncertainty = true, table-format=2.4(4), table-align-uncertainty = true] 
                        S[table-format=1.2]}
\cline{1-8}\cline{1-8} & \textcolor{gray}{$\nu_{\max}$} & \textcolor{gray}{$\Delta\nu$} & \textcolor{gray}{$T_{\text{eff}}$} & \textcolor{gray}{$\text{[Fe/H]}$} &  & &    \\ \hline
{Combination} & {$\alpha$} & {$\beta$} & {$\gamma$} & {$\delta$} & {$\sigma_{\text{sys}}/$R$_\odot$} & {Relative difference} & {$r^2$}  \\ \hline \hline
\rowcolor{aliceblue} 2   &       0.9570   &       -1.955          &       0.6288      &        {--}    &       0.037   &       0.0026(56)      &       0.99 \\
\rowcolor{aliceblue} 4   &       {--}    &       -0.8048         &       2.062   &   0.1378   &       0.16    &       0.000(25)       &       0.99 \\
\rowcolor{aliceblue} 5   &       -0.6593         &       {--}    &       2.953   &   0.2283   &       0.25    &       -0.002(40)      &       0.97 \\\hline
\rowcolor{aliceblue} 6   &       1.008   &       -1.999          &       {--}    &   {--}     &       0.075   &       0.002(13)       &       0.99 \\
\rowcolor{aliceblue} 7   &       {--}    &       -0.7362         &       0.8088      &        {--}    &       0.24    &       -0.001(39)      &       0.98 \\
\rowcolor{aliceblue} 8   &       -0.5591         &       {--}    &       0.7857      &        {--}    &       0.38    &       -0.004(62)      &       0.94 \\\hline
\rowcolor{aliceblue} 9          &       {--}    &       -0.7038         &       {--}         &       {--}    &       0.24    &       -0.000(42)      &       0.97 \\
\rowcolor{aliceblue} 10          &       -0.5353         &       {--}    &       {--}         &       {--}    &       0.36    &       -0.001(62)      &       0.94 \\
\cline{1-8}\cline{1-8}
        \end{tabular}
    \end{threeparttable}
\end{table*}


\section*{Acknowledgements}
The author thanks Amalie Stokholm, J{\o}rgen Christensen-Dalsgaard, Dennis Stello, and the anonymous referee for useful comments and suggestions. 
Funding for the Stellar Astrophysics Centre is provided by The Danish National Research Foundation (Grant agreement no.: DNRF106). 
The numerical results presented in this work were obtained at the Centre for Scientific Computing, Aarhus. \vspace*{-1.2\baselineskip} 

\bibliographystyle{mnras}
\bibliography{Bellinger} 

\appendix\onecolumn
\section{Source code} \label{sec:code}
For the sake of convenience, here is source code in python3 for using the calibrated scaling age relations. 
For the mass and radius relations, please visit this HTTPS URL: \url{https://github.com/earlbellinger/scaling-giants}

\begin{myminted}{scaling\_age.py}
import numpy as np 
from uncertainties import ufloat

# Enter some data for an example red giant whose age we want to estimate 
# Use ufloat(0,0) for any measurement that is not available
nu_max   = ufloat( 140.87,   0.83 ) # muHz, second argument is uncertainty 
Delta_nu = ufloat(  12.715,  0.021) # muHz
Teff     = ufloat(5054,     95    ) # K
Fe_H     = ufloat(  -0.6,    0.1  ) # dex

# Calibrated exponents from Table 2
# P       =       [   alpha,    beta,  gamma,  delta]
P_age = np.array([[- 9.760 , 13.08  , -6.931, 0.4894],  # 1
                  [- 7.778 , 10.77  , -11.05,      0],  # 2
                  [-12.19  , 15.86  ,      0,  1.027],  # 3
                  [       0,  1.396 , -22.32, -1.046],  # 4
                  [  1.084 ,       0, -23.28, -1.165],  # 5
                  [- 8.837 , 11.73  ,      0,      0],  # 6
                  [       0,  0.9727, -14.64,      0],  # 7
                  [  0.6424,       0, -13.82,      0]]) # 8
sigma_sys = np.array([0.25, 0.32, 0.34, 0.82, 0.92, 0.86, 1.2, 1.3])

def scaling_age(nu_max, Delta_nu, Teff, Fe_H, 
        nu_max_ref=104.5, Delta_nu_ref=9.25, Teff_ref=4790, age_ref=4.3):
    # Determine which row of the table to use by checking which entries are 0
    star = np.array([nu_max!=0, Delta_nu!=0, Teff!=0, Fe_H!=0])
    found = False
    for idx, exponents in enumerate(P_age):
        found = np.array_equal(np.nonzero(exponents)[0], np.nonzero(star)[0])
        if found: 
            break 
    
    if not found: # No applicable scaling relation 
        return np.nan 
    
    # Equation 1, plus the systematic error of the corresponding relation 
    alpha, beta, gamma, delta = exponents
    return ((nu_max   /   nu_max_ref) ** alpha  * 
            (Delta_nu / Delta_nu_ref) ** beta   * 
            (Teff     /     Teff_ref) ** gamma  * age_ref * 
            (np.e**Fe_H             ) ** delta) + ufloat(0, sigma_sys[idx])

age = scaling_age(nu_max, Delta_nu, Teff, Fe_H)
print('Age:', '{:.2u}'.format(age), '[Gyr]')
\end{myminted}
\begin{myminted2}
$ python3 scaling_age.py
Age: 7.7+/-1.2 [Gyr]
\end{myminted2}


\label{lastpage}
\end{document}